\documentclass[a4paper,12pt]{article}
\usepackage[pdftex]{graphicx}
\usepackage{caption}
\usepackage{amsmath}
\usepackage{plain} 
\title{Collective memory, consensus, and learning explained by social cohesion}

\author{Jeroen Bruggeman\thanks{Department of Sociology, University of Amsterdam, Nieuwe Achtergracht 166, 1018 WV Amsterdam, the Netherlands. Email: j.p.bruggeman@uva.nl. }}
\date{\today}

\begin{document}
\maketitle



\begin{abstract}
Humans cluster in social groups where they discuss their shared past, problems, and potential solutions; they learn collectively when they repeat activities; they establish social norms; they synchronize when they sing or dance together; and they bond through social cohesion. A group is more cohesive if its members are closer together in their network and are bonded by multiple connections. Network proximity and redundancy are indicated by the second smallest eigenvalue of the Laplacian matrix of the group network, called the \textit{algebraic connectivity}. This eigenvalue is key to explaining and predicting the outcomes of said activities.   
\end{abstract}

Humans live in social groups where they undertake numerous activities \cite{simmel08,coleman88,apicella12}, leading to collective learning \cite{argote90,yelle79}, social cohesion \cite{whiteharary01}, agreement \cite{paluck19}, conventions \cite{judd10,centola15}, collective memories \cite{coman16,momennejad21}, or synchronized movement \cite{jadbabaie04,mcneill95}. Each of these phenomena has been studied separately, based on different models. This paper demonstrates how they are related, thereby enhancing our understanding. To this end, we represent the group networks of the experimental studies of said phenomena with Laplacian matrices, and examine their second smallest eigenvalue, called the \textit{algebraic connectivity}. Laplacians have been used to model diffusion in a wide range of networks \cite{newman18}, and engineers use them to assess the vulnerability of their constructions \cite{moretti19}, but in the social sciences, applications are few and far between, e.g., \cite{bozzo13,shore15b}. Algebraic connectivity of a group increases with redundancy of connections and average network proximity (hops through the net), which are indicative of social cohesion. It turns out that this conception of cohesion explains a wide range of experimental outcomes where information is shared, such as memories \cite{coman16,momennejad21}, individuals' traits \cite{judd10}, conventions \cite{centola15}, opinions \cite{moussaid15,friedkin11,paluck19}, and visual cues of body movement \cite{koul23}. The more cohesive the group, the faster the information is shared. However, groups also have to adapt to their changing social and material environments, for which individuals need some leeway without being suffocated by excessive cohesion. The trade-off between information sharing, leading to homogeneity, and diversity required for adaptability, implies an optimal, rather than maximal, cohesion \cite{ghavasieh24}. 

In this paper, we revisit the experiments, explain their outcomes through algebraic connectivity, and discuss its strengths and weaknesses. The results section has a minimum of mathematical notation, to make it accessible to scholars without a mathematical background. The technicalities are deferred to a methods section, after the discussion.


\section*{Results}
\subsection*{Consensus, collective memories, and synchronization}
\begin{figure}[!ht]
\captionsetup{width=.905\linewidth}
\begin{center}
\includegraphics[width=0.3\textwidth]{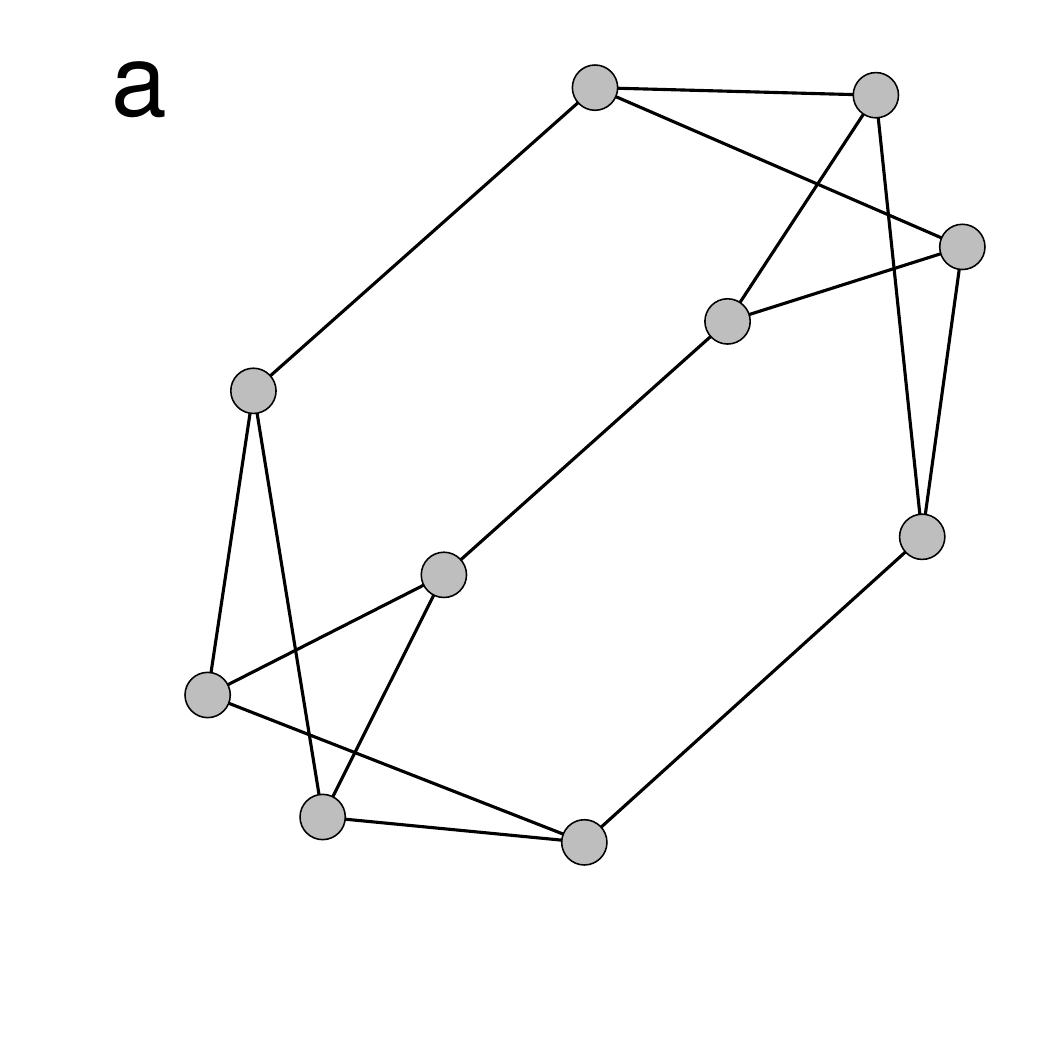}
\includegraphics[width=0.3\textwidth]{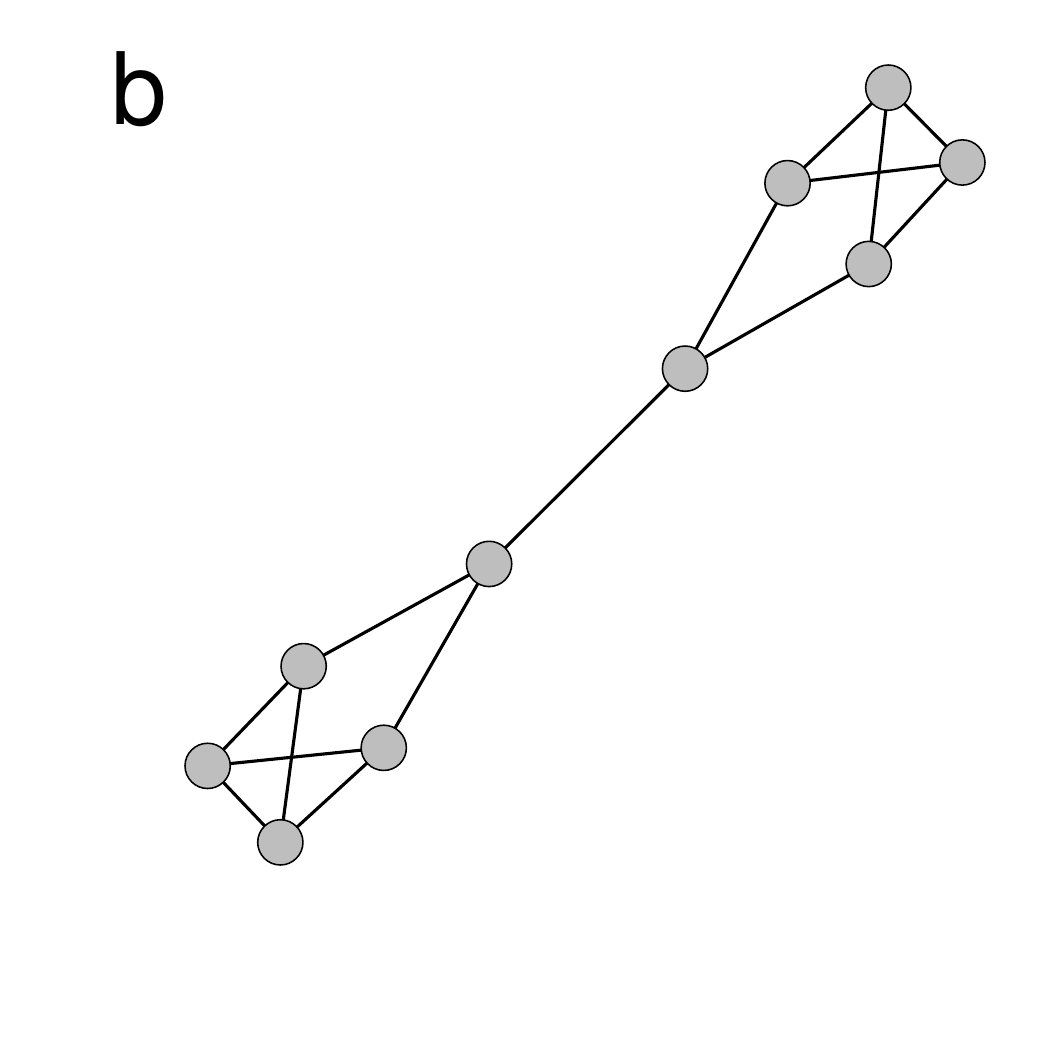} 
\includegraphics[width=0.35\textwidth]{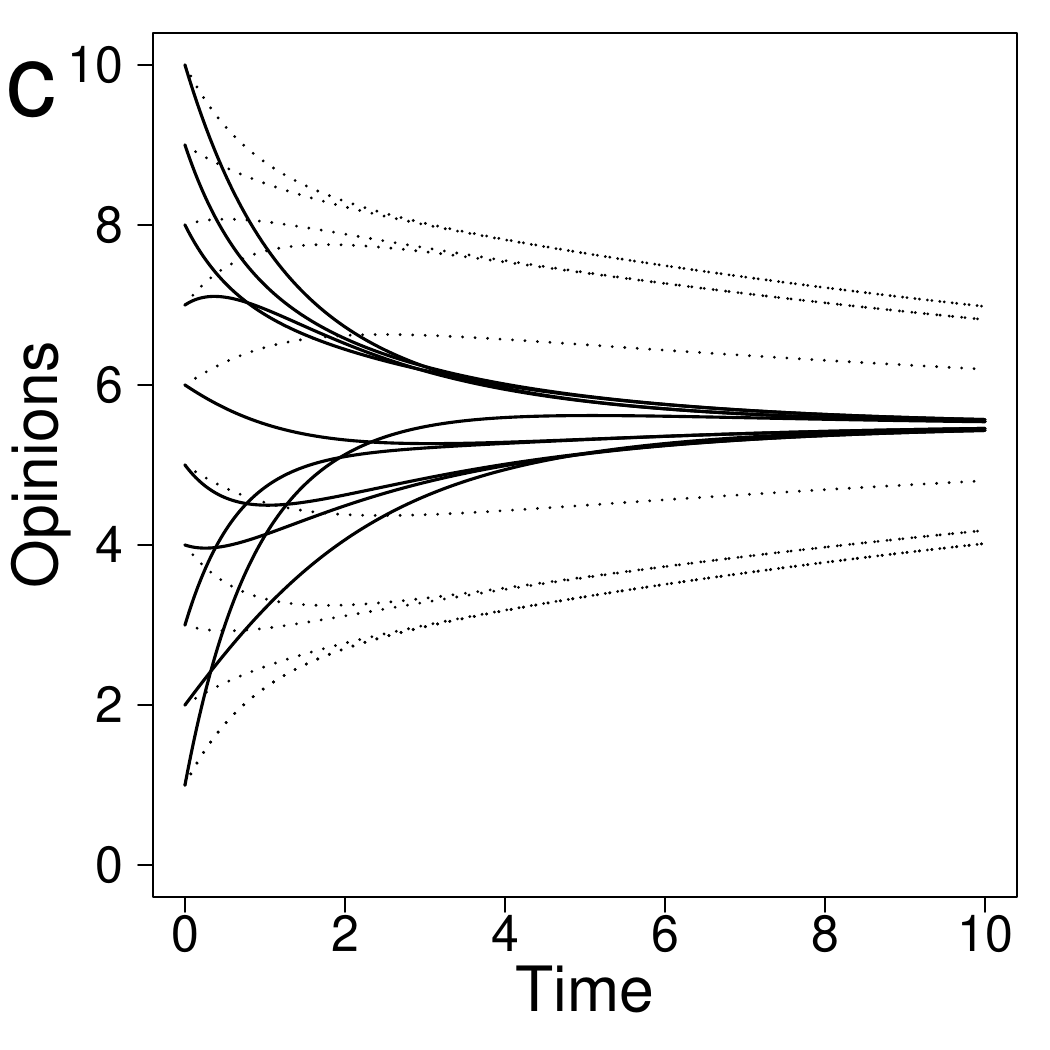}  
\end{center}
\caption{Two networks of the same size and density were used in a mnemonic convergence experiment (adapted from Coman et al.~2016). ({\bf a}) Network with $\lambda_2 = 0.333$. ({\bf b}) Network with $ \lambda_2 = 0.074 $. 
({\bf c}) Memory convergence in network a (continuous lines) and network b (dotted lines) according to the diffusion equation (Eq.~1). }
\label{fig:mnemonic} 
\end{figure}

Information sharing in a group can be modeled as a grand total of pairwise information sharing. When individuals indexed $i$ and $j$ discuss (or notice) each other's positions of a certain kind (attitudes, opinions, behaviors), denoted $y$, their difference is $y_j - y_i$. By using matrix notation, the process of position change, based on these pairwise differences, can be written as 
\begin{equation}
\frac{d\mathbf{y}}{dt} = - \mathbf{L}\mathbf{y},  
\label{eq:influence}
\end{equation}  
where $\bf{L}$ is the network Laplacian (Methods)   
and $\bf{y}$ is a vector of individuals' positions \cite{olfati04}. This equation is also called a diffusion model; over longer time, information reaches individuals $j$ who are further away from sender $i$. It is an efficient notation for well-known and well-replicated social influence models \cite{takacs16,flache17}. The added value over alternative models is in the eigenvalue spectrum of $\mathbf{L}$, i.e., the scalars $\lambda$ that solve $\mathbf{L} \mathbf{y} = \lambda \mathbf{y}$. Initial differences tend to decrease, thus positions $\bf{y}_{0}$ at $t_0$ converge, and the speed of convergence increases with the second-smallest eigenvalue \cite{olfati04}, denoted $\lambda_2$. This so called \emph{algebraic connectivity} \cite{fiedler73} also predicts the synchronization of connected oscillators \cite{jadbabaie04}, for example singing or dancing individuals \cite{mcneill95}, even though models of synchronization differ from the diffusion model \cite{mcgraw07}. Algebraic connectivity is maximal for fully connected networks, zero for networks with disconnected parts, and close to zero if the network is very sparse between dense parts \cite{abreu07}, such as social groups (for example, Fig.~\ref{fig:mnemonic}b). For a given value of $\lambda_2$, convergence lasts longer when initial positions are more diverse. We note that $\lambda_2$ is only loosely related to density, even though adding ties to a network cannot decrease $\lambda_2$ \cite{brouwer11}.
 
The two networks in Fig.~\ref{fig:mnemonic} were used in an experiment \cite{coman16} that drew considerable attention \cite{spinney17}. The participants conversed about their past (i.e., a story made up by the researchers that subjects first had to memorize) for a certain amount of time. 
The authors concluded that in the network with more mnemonic alignment, in Fig.~\ref{fig:mnemonic}a, clustering is lower and the mean distance is smaller than that in Fig.~\ref{fig:mnemonic}b. The diffusion model (Eq.~\ref{eq:influence}) explains the two outcomes by providing the mechanism; it shows how the outcome results from interacting individuals in their networks; see Fig.~\ref{fig:mnemonic}c. The model also makes it possible to relate specific findings, such as this one, to numerous others, comprising a variation of the memory experiment (Appendix). 

Another experiment involved the formation of a convention among multiple alternatives proposed by the participants \cite{centola15}. Subjects interacted over 25 rounds, communicating with one alter in each round. Three networks were compared, each with $n = 24$: a circular lattice where everyone had four ties ($\lambda_2 = 0.084$); a random network where everyone had four ties on average ($\lambda_2 = 0.256$); and a network where subjects were connected to a next alter in every subsequent round, homogeneously spread across the network. In the latter network, this procedure resulted in a clique (fully connected network) at round 23 ($\lambda_2 = 1.043$), when everyone had interacted with everyone else. In the two networks with low connectivity (lattice and random), a group convention was never achieved in time, in contrast to the clique, with the highest connectivity, where it was always achieved.

\subsection*{Cohesion}
Underlying the experimental outcomes is the social cohesion of the network, which is the network structure that bonds a group together and facilitates information transmission \cite{moodywhite03}. Because information transmitted over longer distances tends to deteriorate \cite{moussaid15,eriksson12} or not arrives at all \cite{dodds03}, it is important for group members that distances to sources are short, and there is network redundancy such that a noisy or biased message through one path (concatenation of ties) can be corrected by messages through other, nonoverlapping, paths \cite{whiteharary01}, as in Fig.~\ref{fig:mnemonic}a compared to b. Redundancy is also at the heart of complex contagions, where $j$ is more likely to be won over by $i$ when $j$ receives $i$'s information through multiple channels rather than one \cite{centola07}.
Algebraic connectivity is larger in networks with shorter distances and a larger minimum of node-independent paths connecting arbitrary pairs of nodes (of which there are 3 in Fig.~\ref{fig:mnemonic}a and 1 in b). It is also larger when chords cross-connect these paths, an aspect of redundancy that has been largely overlooked in the literature, whereas they turned out to be crucial for reliable transmission in an experiment with long paths \cite{eriksson12}. Redundancy and proximity increase the robustness of a network, yield rapid consensus on important issues, and intensify interaction rituals \cite{collins04}, such as synchronous singing and dancing \cite{reddish13}, which in their turn contribute to the sensed bonding of a group, or solidarity, and facilitate collective action \cite{bruggeman24}. 

Of these network traits and their consequences, people are dimly aware at best. Social cohesion is an emergent phenomenon resulting from individuals establishing and breaking social contacts, mostly in their local environment (when leaving out social media algorithms). These local decisions result in relatively dense groups that are sparsely interconnected \cite{newman18}, with relatively weak intergroup cohesion. We note that collective action of large networks also depends on shared social norms, leadership, and sometimes ideology or religion, which are not indicated by $\lambda_2$. Therefore, $\lambda_2$ tells us less about the collective action potential of large networks than of smaller groups, or clusters, embedded in these networks. 

\subsection*{Learning}
In more cohesive groups (if not too large), not only does convergence to consensus proceed faster but also the collective production of goods, such as building a house. The time, $t$, it takes to complete a collective task is described by the learning curve \cite{yelle79}: $t$ decreases at a decreasing rate with the number of times the activity has been executed in the past. Due to task repetition, individuals learn, but we are interested in the group level effect: the group network becomes more efficient through individuals finding shortcuts in their social ``problem space" \cite{newell72} with increasing probability \cite{shrager88}, connecting to colleagues at closer network distances. It has been shown computationally (a.o.~in small world models) that probabilistic tie relaying decreases mean distance \cite{watts98}, $\overline{\mathcal{D}}$, and in a mathematical model, the relation between $\overline{\mathcal{D}}$ and $t$ (keeping density fixed) was identical to the learning curve \cite{shrager88,huberman01}. These scholars did not examine path redundancy, $\mathcal{K}$, but because $\lambda_2$ increases with decreasing distance (Methods), we conjecture that $\lambda_2$ has a learning curve relation with $t$.  

\begin{figure}[!ht]
\captionsetup{width=.905\linewidth}
\begin{center}
\includegraphics[width=0.45\textwidth]{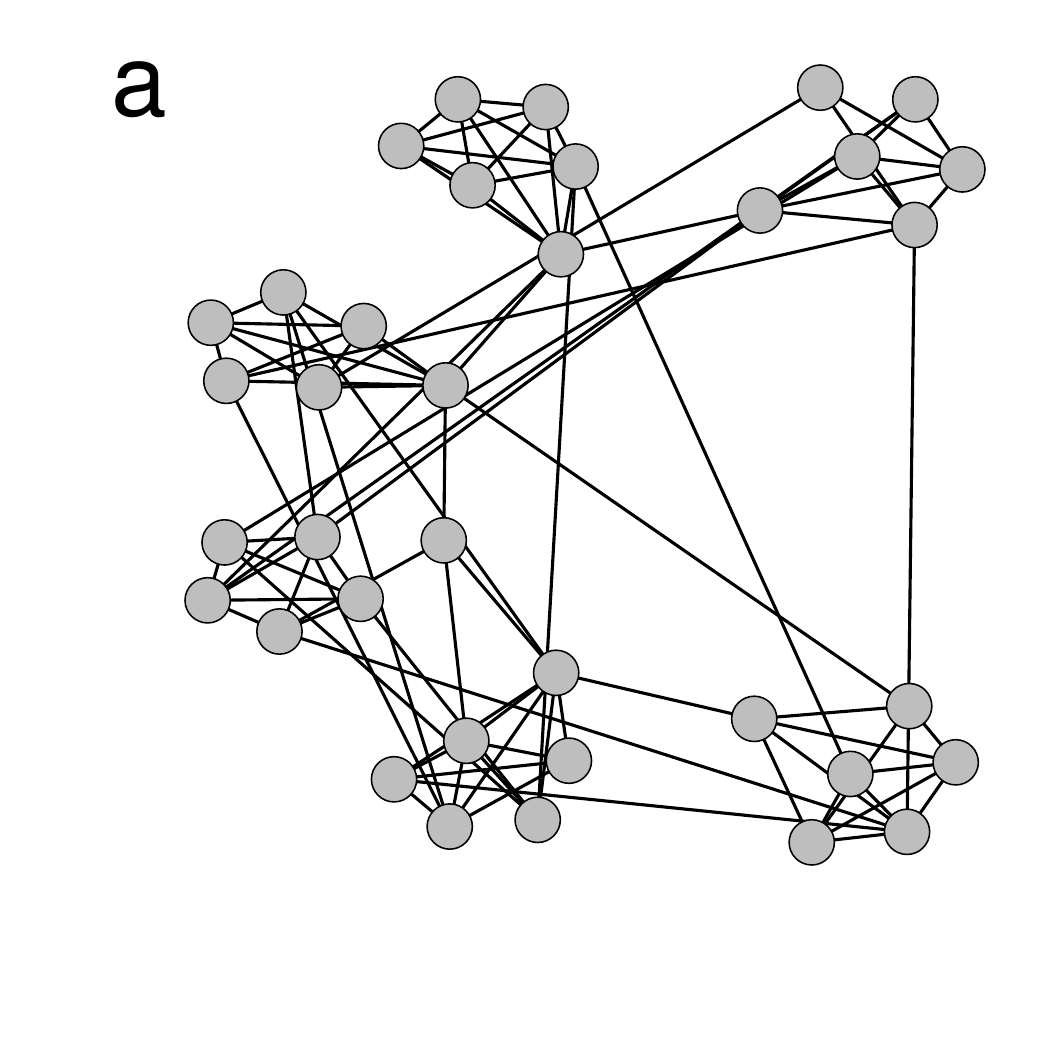} 
\includegraphics[width=0.45\textwidth]{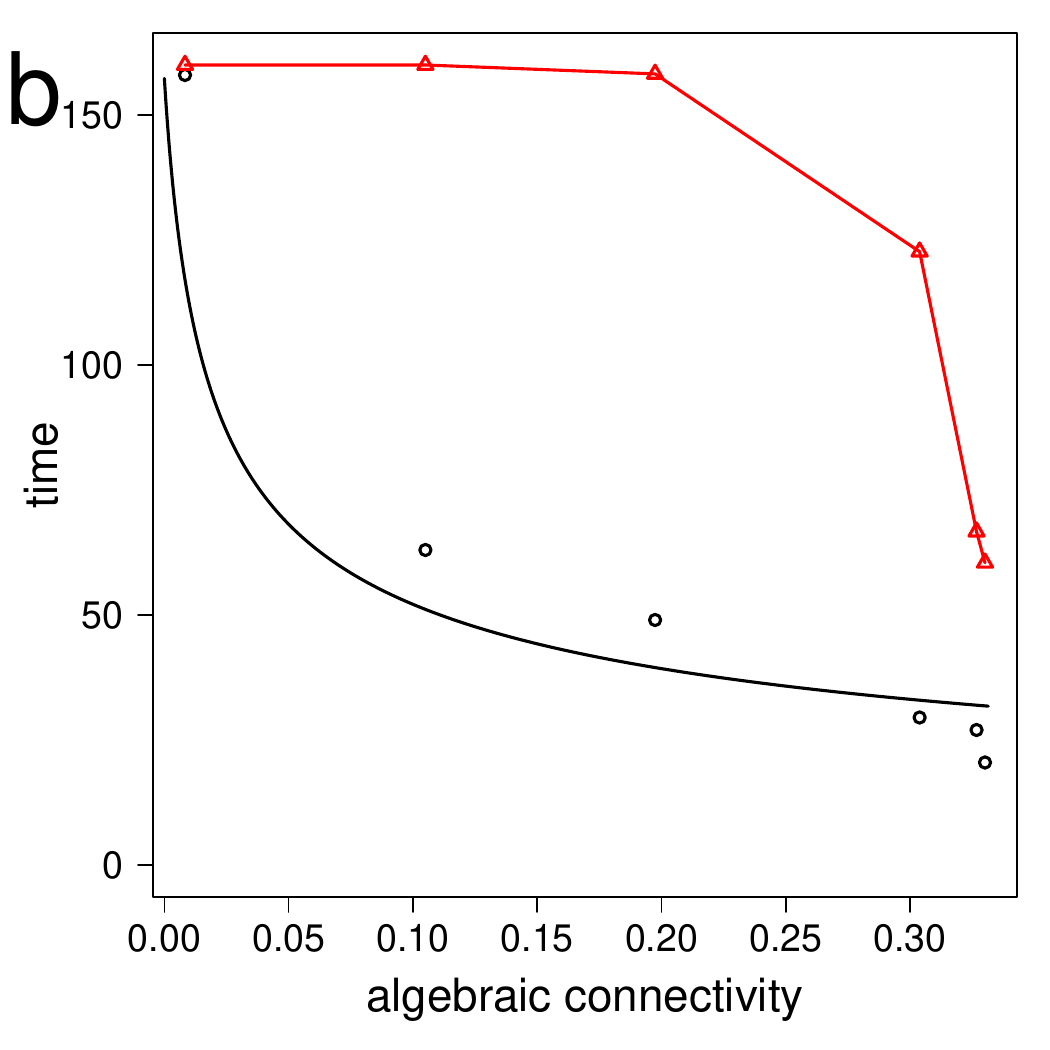} 
\end{center}
\caption{Node coloring experiment. (a) Network with six clusters and a tie-rewiring probability of $p = 0.1$ (adapted from Kearns et al.~2010). (b) Time (in seconds) to reach consensus as a function of the algebraic connectivity of the network. Note that the data (circles) was obtained for given chances of random tie relaying, but here, pertinent levels of algebraic connectivity are plotted.  The predictions from the original myopic heuristic model (in seconds) are shown in triangles.}
\label{fig:kearnsnet} 
\end{figure}

To find out if this is true, we revisited a network experiment wherein every subject was randomly assigned an initial color, and was instructed to synchronize their color with their network neighbors \cite{judd10}. (The number of colors to choose from was the chromatic number.) The time it took to achieve one overall color (i.e., the collective task) was measured in six networks, all with the same density and 36 subjects. The first had six densely connected clusters with one tie connecting each cluster to the next; the remaining five networks were obtained by randomly rewiring the ties of the first network with increasing probability, $p \in \{0,0.1,0.2,0.4,0.6,1\}$, while keeping the, progressively sparser, clusters connected to each other (Table~\ref{tab:coloring}). The network with $p = 0.1$ is illustrated in Fig.~\ref{fig:kearnsnet}a. 

\begin{table}[!ht]
\begin{center}
\begin{tabular}{|l|l|llllll|}
\hline
treatments & $p$ & 0 & 0.1 & 0.2 & 0.4 & 0.6 & 1 \\ \hline
outcome & $t$ & 158 & 63 & 49 & 29.5 & 27 & 20.5 \\ \hline
predictors & myop & 180 & 180 & 178 & 138 & 75 & 68 \\ 
& $\lambda_2$ & 0.0083 & 0.1050 & 0.1974 &  0.3038  &  0.3267  &  0.3301    \\
&  $\overline{\mathcal{D}}$  & 3.5714 & 2.5901 & 2.3707 & 2.2502 & 2.2305 & 2.2229 \\ 
& $\mathcal{K}$ & 1 & 2.5220 & 2.4734 & 1.8558 & 1.5726 & 1.5188  \\      
\hline
\end{tabular}
\caption{The network node coloring experiment. From top to bottom: rewiring probability, $p$; time to task completion in seconds, $t$; and the predictions of the myopic decision model, myop, in seconds. All three are from the original paper. Then, there is algebraic connectivity, $\lambda_2$; mean distance, $\overline{\mathcal{D}}$; and the minimum number of node-independent paths connecting arbitrary pairs of nodes, $\mathcal{K}$. The latter three were averaged over 1000 simulated networks.}
\label{tab:coloring}
\end{center}
\end{table}

In contrast to teams and business companies \cite{argote90}, where individuals find shortcuts in their network by themselves, the researchers implemented the shortcuts in the coloring experiment. Apart from this difference, the sequence of treatments can be regarded as networkwise identical to a sequence of snapshots of a learning process. Bearing this in mind, a learning curve was fitted to the data from the coloring experiment, substituting $\lambda_2$ for the number of task repetitions (Fig.~\ref{fig:kearnsnet}b). The functional form most often used is a power law, $t = a \lambda_2^{-b}$ (and $b > 0$), and in our case, $b = 0.434$; this is close to the average in industry, $b \approx 1/3$ \cite{argote90}. Because the authors of the coloring experiment did not share their data, statistics of fit are not reported. For each $p$ value, $\lambda_2$ was calculated as the mean of 1000 simulated networks.
Time $t$ was estimated by applying a measuring rod to an enlarged print of Fig.~4 from the original paper. For the latter, the inter-rater correlations of the author with two colleagues were 0.9997 and 0.9996; the reported times (Table~\ref{tab:coloring}) are the means of our three measurements. Despite the uncertainty of the experimental networks, it is clear that the learning curve predicts the experimental outcomes much better than does the authors' myopic decision model with a concave, rather than convex, curve (Fig.~\ref{fig:kearnsnet}b). 

As a simpler explanation of the learning curve, mean distance will not do. The graph of $\overline{\mathcal{D}}$ with $t$ is almost a straight line (Fig~\ref{fig:distance2}a), not a learning curve, and when $p$ (driving $\overline{\mathcal{D}}$) increases, $\mathcal{K}$ changes, too. $\mathcal{K}$ has a non-monotonic relation with $t$ (Table~\ref{tab:coloring}) that should be controlled for. When, accordingly, using both proximity and redundancy to explain how network structure improves work efficiency, $t$, the most parsimonious way is through algebraic connectivity.   

\section*{Discussion}
An important lesson from the coloring experiment is that the incremental effect size of algebraic connectivity is large at low levels, whereas further increasing connectivity to higher levels has small effects. In other words, to gain most of the advantages of cohesion, a modest level suffices. This aligns with the findings of other studies, in which the effect of messages through multiple ties was examined. 
In one of these studies, the chance to win subjects over to accept a health improving behavior increased with the number of ties through which they received health messages, but at a decreasing rate, and more than three ties had no significant effect above three ties \cite{centola10}. The only exception seemed to be the presence of a time limit, when a group result had to be achieved in a short period of time (in the convention experiment, \cite{centola15}), but the clique network was the result of our temporal densification of the experimental network; at each moment, the subjects interacted with one alter, not with all of them. 

Along with diminishing advantages, there are increasing disadvantages. If people too easily converge to the majority, the result is not necessarily wisdom of the crowd but oftentimes a form of groupthink: a commitment to an inferior solution to a given problem \cite{janis72}, a lack of innovation due to cultural homogeneity \cite{page07}, a memory of events that never happened \cite{coman16}, the adoption of a conspiracy theory \cite{greve22}, or the punishment of critics with valuable feedback \cite{glazer89}. For information from multiple sources to be advantageous, two conditions must be fulfilled. First, knowledgeable people should not let themselves be influenced by sources with inferior information (\cite{bruggeman17,centola23}). Second, the least knowledgeable individuals should either be aware that they can learn from their average social environment \cite{centola23} or specifically know which sources are best, and relay their ties accordingly \cite{bruggeman17}. Both conditions are easier to meet at modest levels of algebraic connectivity (and density) when the network is not too efficient and social pressure is not too strong.

In sum, most of the benefits of algebraic connectivity can be had at relatively low levels (depending on $n$), whereas further increasing connectivity by means of extra ties entails costly tie maintenance, increasing information overload, and higher risks of groupthink, while rarely yielding additional benefits. A challenge for future research is to determine the sweet spot. A promising development into this direction is the diffusion model applied to the trade-off between the network's ability to adapt, which requires diverse behavior across individuals (instead of homogeneity), and its information transmission capability, which benefits from strong connectivity \cite{ghavasieh24}. Currently, it is unknown if this trade-off is the most important one in social life, which is for future research to find out.

In this paper, we have shown that the diffusion model with its algebraic connectivity predicts the outcomes of various processes that converge. 
The range of applications turns out to be more general than expected, comprising categorical (versus continuous, Eq.~\ref{eq:influence}) positions, and asynchronous (versus simultaneous) interactions.
However, the model cannot predict divergence, for example choosing a color differing from one's neighbors \cite{kearns06} or the division of labor in general. Nor can it predict when a discussion about a controversial issue leads to polarization instead of convergence \cite{baumann20,santos21a,altafini13}.  Furthermore, there are challenges to social groups beyond convergence/divergence, for example, finding solutions to complex problems \cite{shore15}, for which algebraic connectivity has limited use. 
Nonetheless, a Laplacian network representation of social groups interconnects numerous findings from multiple fields and improves their explanations.
It is also practical; if your meetings last too long, increase their algebraic connectivity! 

\section*{Methods}
\subsection*{Laplacian matrix}
In a set of $n$ individuals, indexed $i$ and $j$, their social ties are denoted $a_{ij} \geq 0$, which in turn are elements of adjacency matrix $\mathbf{A}$.  If $i$ pays attention to $j$, tie $a_{ij} > 0$, and the social influence of all others on $i$'s position on issue $y$ is written as the sum total of pairwise differences \cite{flache17}, $\sum_{j=1}^n a_{ij}(y_j - y_i)$. To write this for everyone concisely (Eq.~\ref{eq:influence}), a vector $\mathbf{y}$ of initial positions is used as well as a Laplacian matrix
\begin{equation}
\mathbf{L} =  \mathbf{D} - \mathbf{A},  
\label{eq:laplace}
\end{equation}  
with $\mathbf{A}$ the adjacency matrix of the network \cite{chung97}.  $\mathbf{D}$ is a diagonal matrix with $d_{ii} = \sum_j a_{ij}$ and $d_{ij} = 0$ if $i \neq j$. The definition of $\mathbf{L}$ implies that $\mathbf{L}\mathbf{1} = 0$, hence its smallest eigenvalue is $\lambda_1 = 0$. The number of components (i.e., from every node, there is a path to every other node) is indicated by the number of eigenvalues equal to zero. This paper is about networks that are one component. 

Results are most straightforward if $\mathbf{A}$ is binary and symmetric ($a_{ij} = 1$ or $a_{ij} = 0$, and $a_{ij} = a_{ji} $); however, this yields the illusions that larger groups with the same density have stronger cohesion and reach consensus more quickly, which are both false. To sidestep these incorrect ideas, the row-normalized adjacency matrix $\bf{W}$ can be used instead of $\mathbf{A}$, with $w_{ij} = a_{ij}/ \sum_{j=1}^n a_{ij}$ such that $\sum_j w_{ij} = 1$, in line with social influence models in general \cite{friedkin11}. Then, $d_{ii} = 1$ for all individuals who pay attention to at least one social contact. Accordingly, $\mathcal{L} =  \mathbf{D} - \mathbf{W}$ was used in the examples above. If $w_{ij} \neq w_{ji}$, this poses no problem in general; however, one-directional ties ($w_{ij} > 0$ and $w_{ji} = 0$) may entail complex eigenvalues without meaningful interpretation in social life, and hardly contribute to social cohesion. Therefore one might consider (depending on the research question) to remove these ties from the data before measuring $\lambda_2$. 

In software packages (e.g., igraph), one finds the normalized Laplacian, $\mathbf{L}_{nor} = \mathbf{D}^{-1/2} (\mathbf{D} - \mathbf{A}) \mathbf{D}^{-1/2}$, not the row-normalized Laplacian. Although $\mathbf{L}_{nor} \neq \mathcal{L}$, it is convenient that $\lambda_2(\mathbf{L}_{nor}) = \lambda_2(\mathcal{L}).$

Whichever of the three Laplacians is used, when $d\mathbf{y}/dt = - \mathbf{L}\mathbf{y}$ (Eq.~\ref{eq:influence}) is integrated, the solution, $\mathbf{y}_t = e^{-\mathbf{L}t}\mathbf{y}_0$, is unpacked in terms of the eigenvectors, $\mathbf{v}_i$, and eigenvalues of $\mathbf{L}$ \cite{sayama15}. 
Because $\mathbf{y}_0 = b_1\mathbf{v}_1 + b_2\mathbf{v}_2 + ... + b_n\mathbf{v}_n$,
\begin{equation}
\mathbf{y}_t = \sum_{k = 1}^n b_ke^{-\lambda_k t}\mathbf{v}_k. 
\label{eq:integration_solution}
\end{equation} 
The outcome is dominated by the smallest eigenvalue larger than zero, $\lambda_2$. Moreover, because $e^{-\mathbf{L}t} = \sum_{k = 0}^{\infty} \frac{(-t)^k}{k!}\mathbf{L}^k$,  information remains local if $t$ is small (i.e., long paths, at large $k$, become negligible), but diffuses further for larger $t$ \cite{ghavasieh24}. 

When there is variation across individuals' susceptibilities to social influence, Eq.~\ref{eq:influence} can be expanded to 
\begin{equation}
\frac{d\bf{y}}{dt} = - \bf{S}\mathcal{L}\bf{y}, 
\label{eq:laplace2}
\end{equation}  
with diagonal matrix $\bf{S}$ = $diag(s_{ii})$, $s_{ii} \neq s_{jj}$, and for all $i \neq j$, $s_{ij} = 0$. The larger the variation in susceptibility is, the less accurate algebraic connectivity is at predicting the speed of convergence, which does not impede numerically solving Eq.~\ref{eq:laplace2}. When susceptibility varies, stubborn individuals (with low $s_{ii}$) have greater influence on $\mathbf{y}_t$. This is also true for leaders to whom others direct their strongest ties.

Approximating the optimal spectrum (including the algebraic connectivity) of a network with $n$ nodes, growing (hypothetically) from zero ties to a given number of ties \cite{ghavasieh24}, goes beyond this paper.  Very briefly, a statistical propagator, $\mathcal{U}_t =  e^{-\mathbf{L}t}/n$ (where $2t$ is rescaled to $t$), is divided by a partition function, $Z = -\mathrm{Tr}(\mathcal{U}_t)$, yielding a network density matrix, $\boldsymbol\rho_t$. Implementing $\boldsymbol\rho_t$ in the von Neumann entropy, 
$\delta = -\mathrm{Tr}(\boldsymbol\rho_t log \boldsymbol\rho_t)$, yields a measure of response to perturbations, and $F = -log(Z)/t$ is a measure of communication speed. $\mathbf{L}$ thus plays a key role in both. Defining $Q = \partial \delta/t$ and $V = \partial F$, the trade-off characterizing the network formation process is $\eta = 1 - |Q|/V$. ``The network with higher $\eta$ better balances the loss in response diversity with the gain in information flow" \cite{ghavasieh24}.

\subsection*{Algebraic connectivity and social cohesion}
In the field of mathematics, the majority of results are based on symmetric binary matrices \cite{vanmieghem10}, instead of row-normalized matrices, including the results in this section. The presentation follows the order of the main text: (\textit{i}) proximity,  (\textit{i}) path redundancy, and (\textit{iii}) chords.

\begin{figure}[!ht]
\captionsetup{width=.905\linewidth}
\begin{center}
\includegraphics[width=0.45\textwidth]{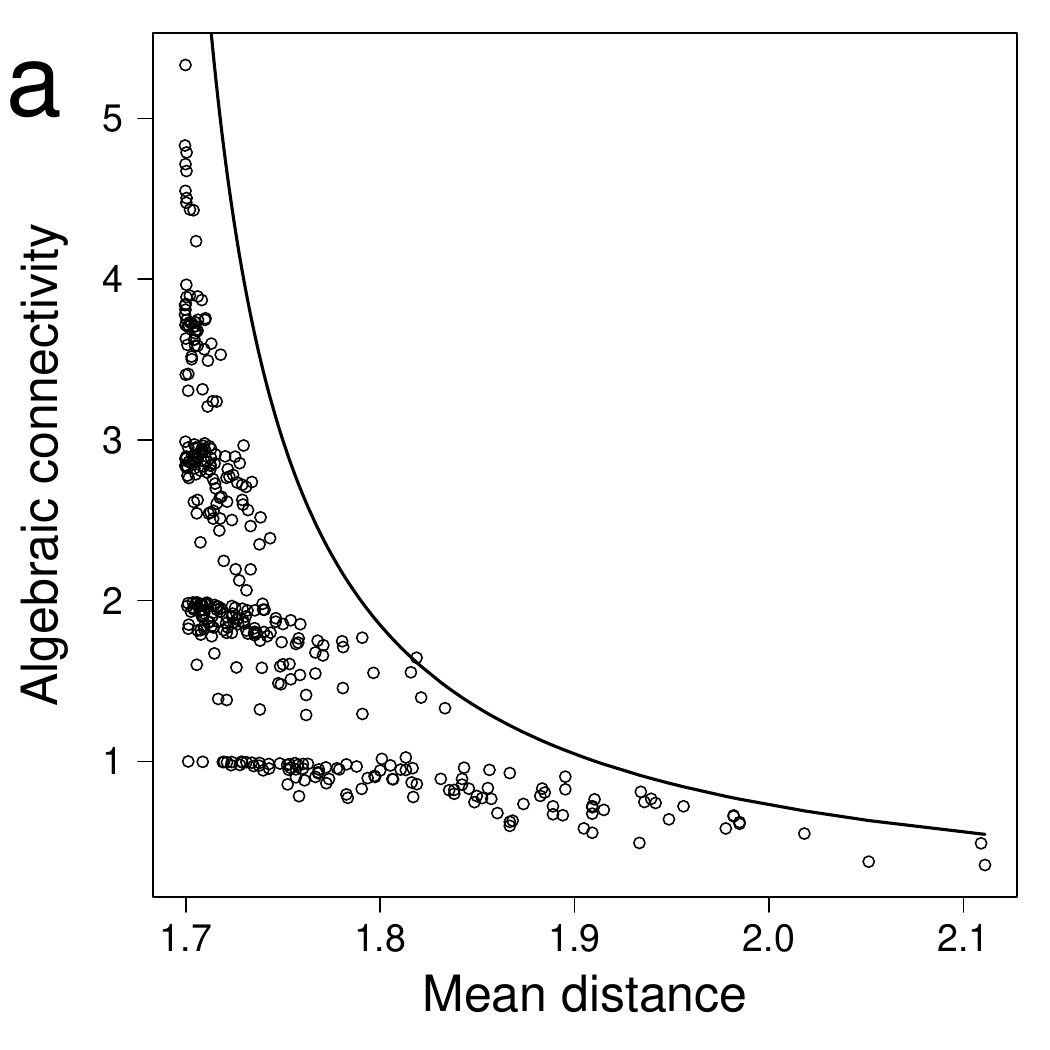}  
\includegraphics[width=0.45\textwidth]{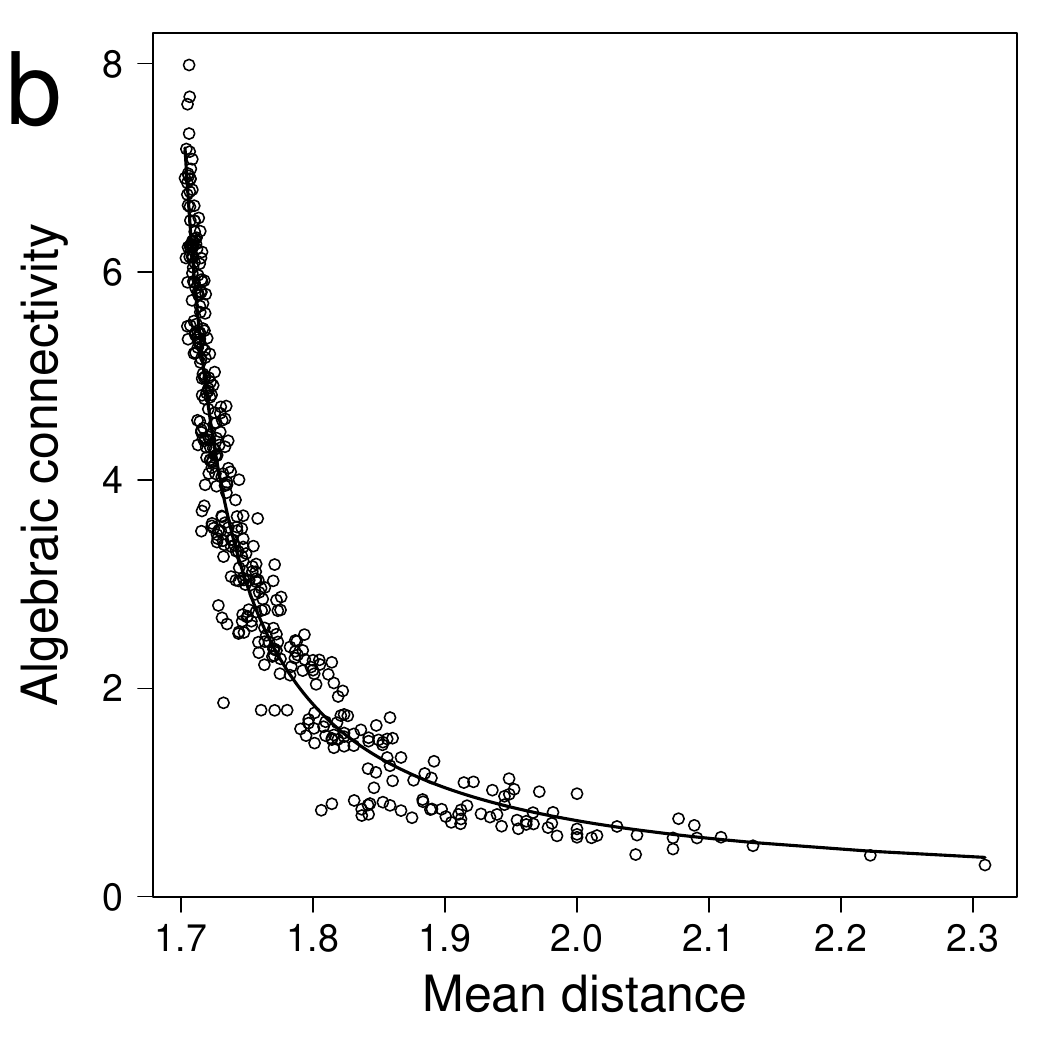}
\end{center}
\caption{Algebraic connectivity is bounded by mean distance. ({\bf a}) Random graphs with skewed degree distributions (size range $[10, 50]$; density = 0.3) and the upper bound of $\lambda_2$, based on Eq.~\ref{eq:distance2}. ({\bf b}) Random graphs with Poisson degree distributions (same sizes and density as in a) with Eq.~\ref{eq:distance2} fitted to the data.}
\label{fig:distance} 
\end{figure}

\begin{figure}[!ht]
\captionsetup{width=.905\linewidth}
\begin{center}  
\includegraphics[width=0.45\textwidth]{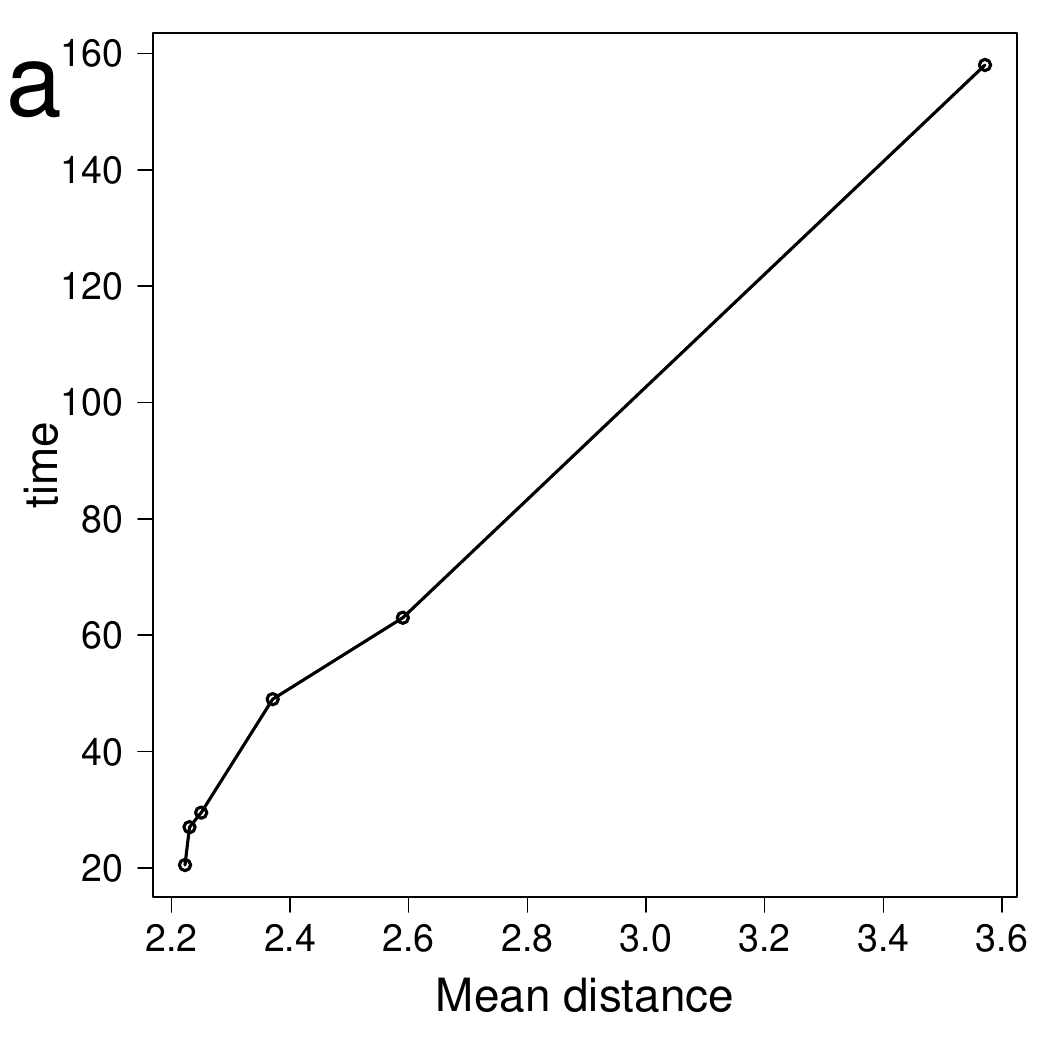}
\includegraphics[width=0.45\textwidth]{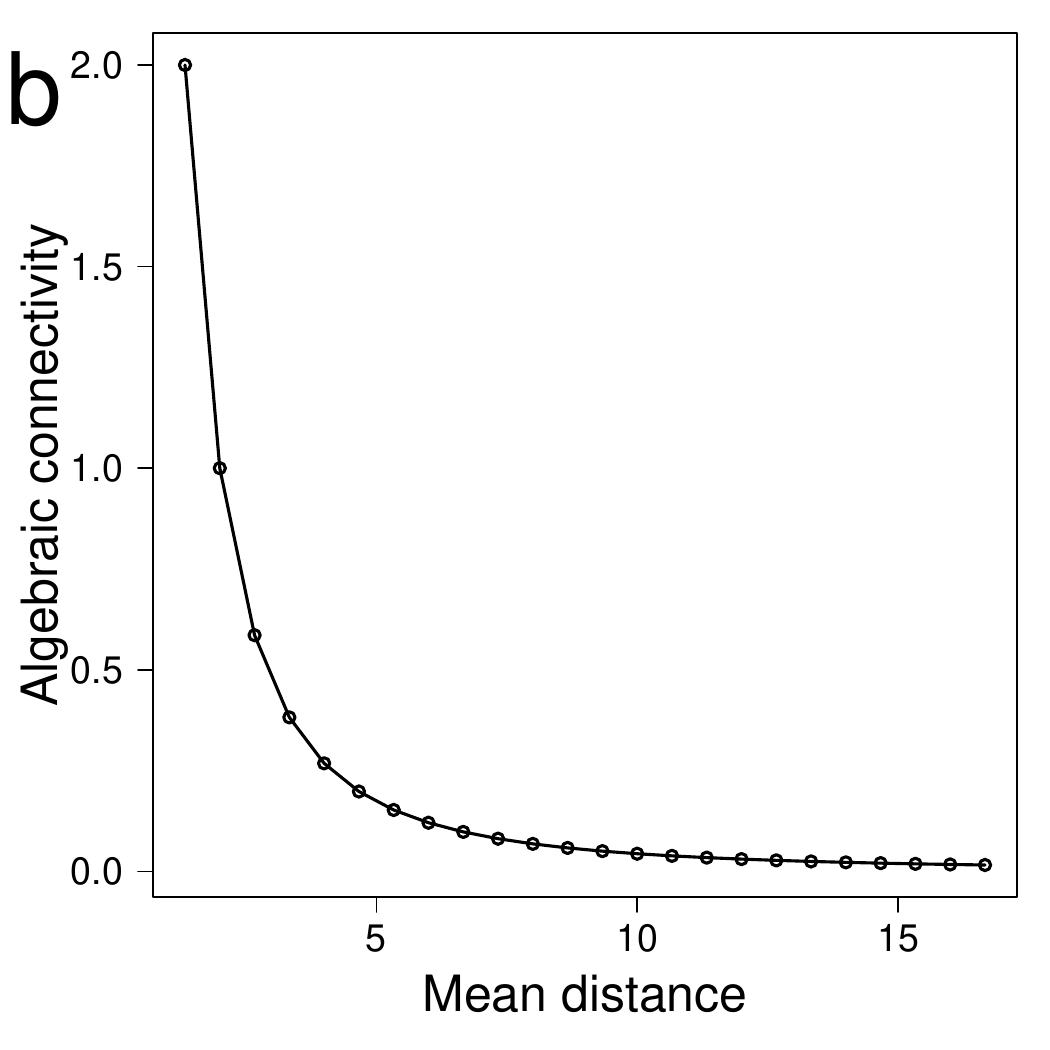}
\includegraphics[width=0.45\textwidth]{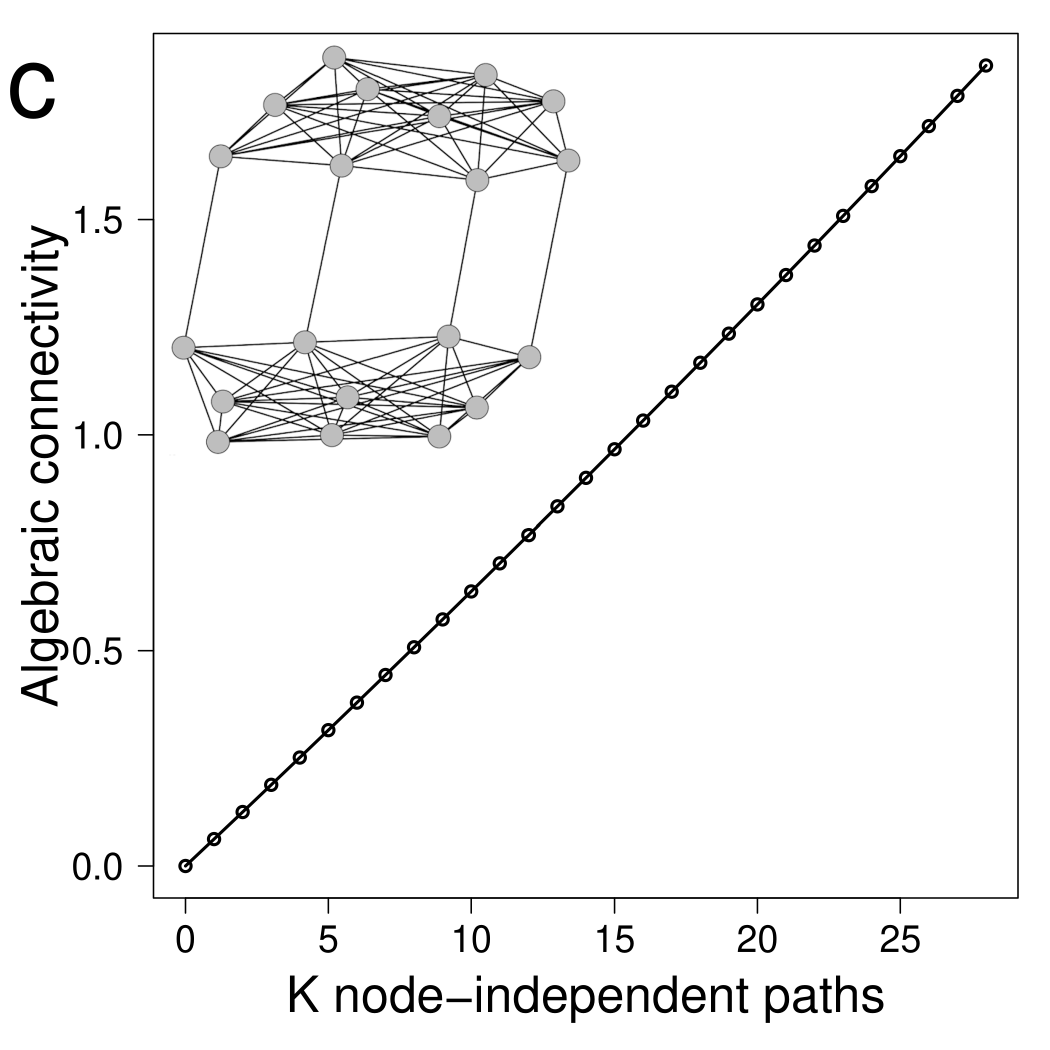}
\includegraphics[width=0.45\textwidth]{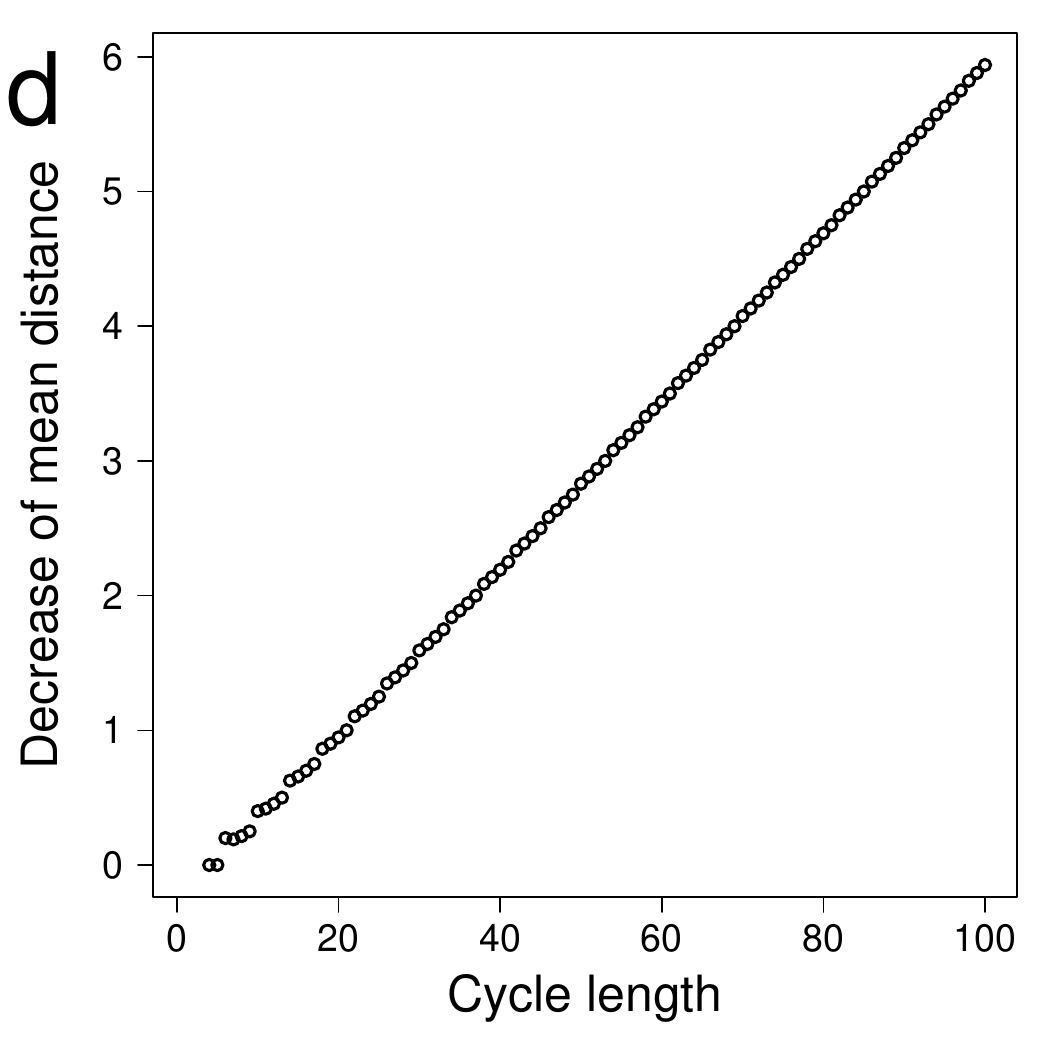}
\end{center}
\caption{Network distance and connectivity.  ({\bf a}) Time to task completion with mean distance in the node coloring experiment. ({\bf b}) Algebraic connectivity with mean distance for square lattice networks (length range [2, 25]). Compare with the random networks in Fig.~3 (main text). ({\bf c}) Algebraic connectivity with node-independent connections between two groups. Initially, the two groups were separate cliques (each, $n = 30$); subsequently, independent connections were placed between them while making them sparser to keep density constant (inset). ({\bf d}) For chordless cycles with a length $l \geq 6$, cross-connecting them midway (or almost midway, when the number of nodes is odd) decreases their mean distance by a magnitude shown on the vertical axis. }
\label{fig:distance2} 
\end{figure}

\textit{Proximity}. The expected, or average, path length, $\overline{\mathcal{D}}$, imposes an upper bound on algebraic connectivity
\begin{equation}
\lambda_2 \geq \frac{2}{(n-1)\overline{\mathcal{D}}-0.5(n-2)}, 
\label{eq:distance}
\end{equation}  
and so does $\mathcal{D}_{max}$, the diameter, $\lambda_2 \geq 4/(n\mathcal{D}_{max})$. Van Mieghem \cite{vanmieghem10} provided proofs. The upper bound is illustrated for a range of networks with a skewed degree distribution but otherwise random (size range [10, 50]; density = 0.3; Fig.~\ref{fig:distance}a). If the degree distribution is Poisson instead of skewed (same size range and density; Fig.~\ref{fig:distance}b), the data points are close to the curve
\begin{equation}
\lambda_2 = \frac{c_1}{\overline{\mathcal{D}}+ c_2}.
\label{eq:distance2}
\end{equation} 
When, for example in a square lattice, one cancels out random fluctuations of redundancy across simulated networks (as in Fig.~\ref{fig:distance}b), the data points are exactly on the curve (Fig.~\ref{fig:distance2}b), even though density cannot be kept constant; it decreases with size.  

\textit{Path redundancy}. For all networks that are no cliques (i.e., are at least one tie short), $\lambda_2 \leq \mathcal{K}$, where $\mathcal{K}$ is the minimum number of nodes that have to be removed to make the network fall apart \cite{fiedler73}. $\mathcal{K}$ also equals the minimum number of non-overlapping (i.e., node-independent) paths connecting arbitrary pairs of nodes \cite{menger27,whiteharary01}, for which $\lambda_2$ thus imposes a lower bound. Furthermore, $\lambda_2 \leq k_{min} $, where $ k_{min}$ is the smallest degree in the network \cite{vanmieghem10}. Computationally, we find a linear relation. When starting out with two separate cliques ($\lambda_2 = \mathcal{K} = 0$), and progressively cross connect them with node-independent ties while removing ties within the two groups to control for density, we find $\lambda_2 = c \mathcal{K}$ (Fig.~\ref{fig:distance2}c). Inevitably, the mean distance slightly decreased during this procedure, from 2 (when the first connection between the cliques was made) to 1.5; if it were held constant, there should be a slight curvature.

\begin{figure}[!ht]
\captionsetup{width=.905\linewidth}
\begin{center}
\includegraphics[width=0.3\textwidth]{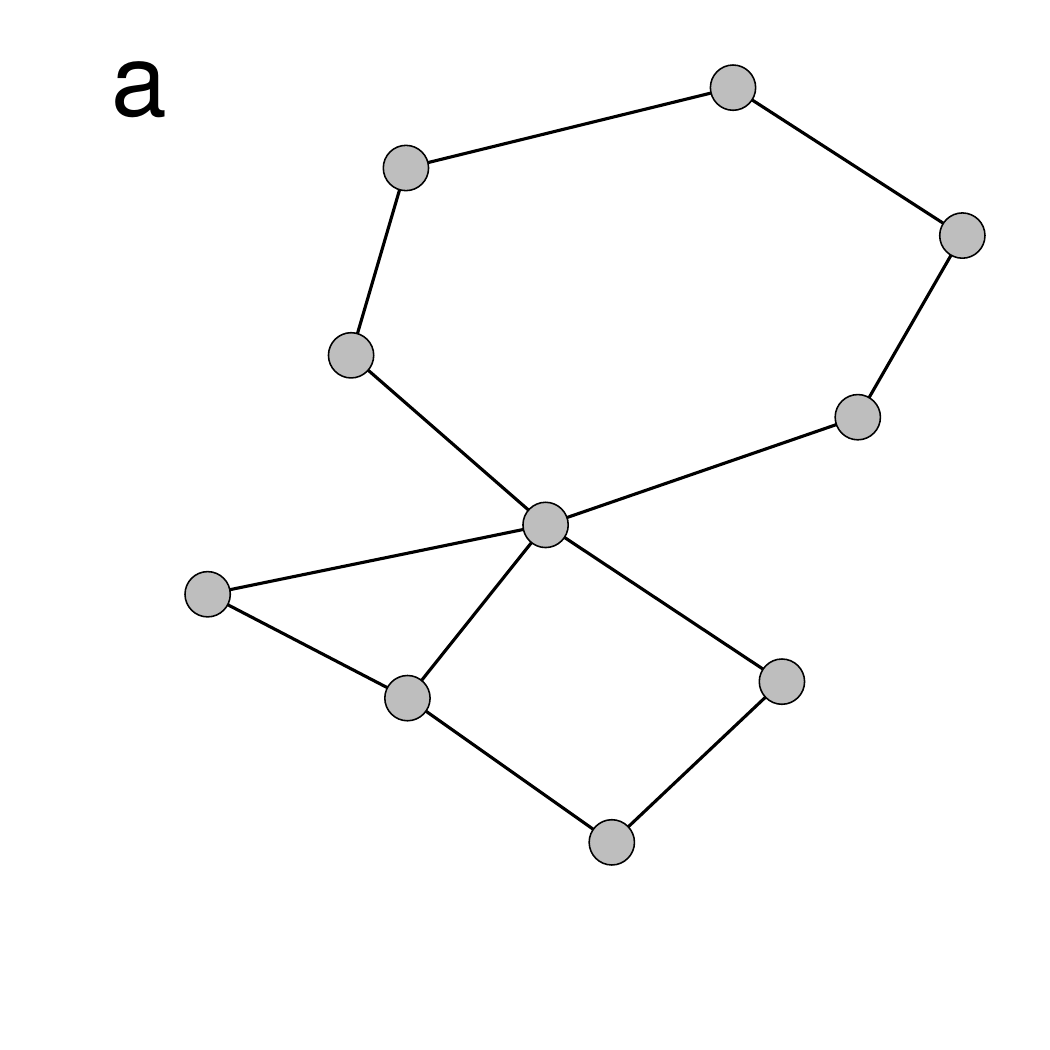}
\includegraphics[width=0.3\textwidth]{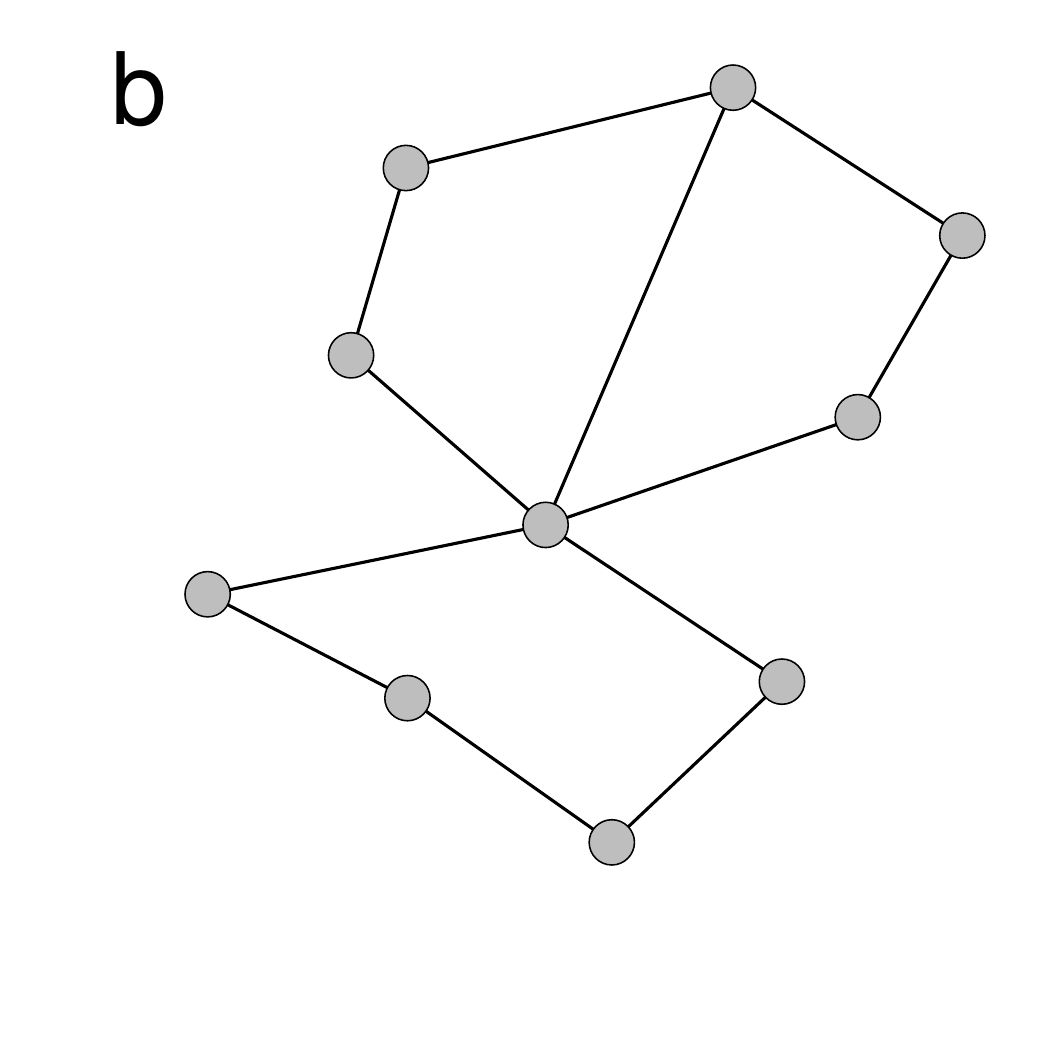} 
\includegraphics[width=0.3\textwidth]{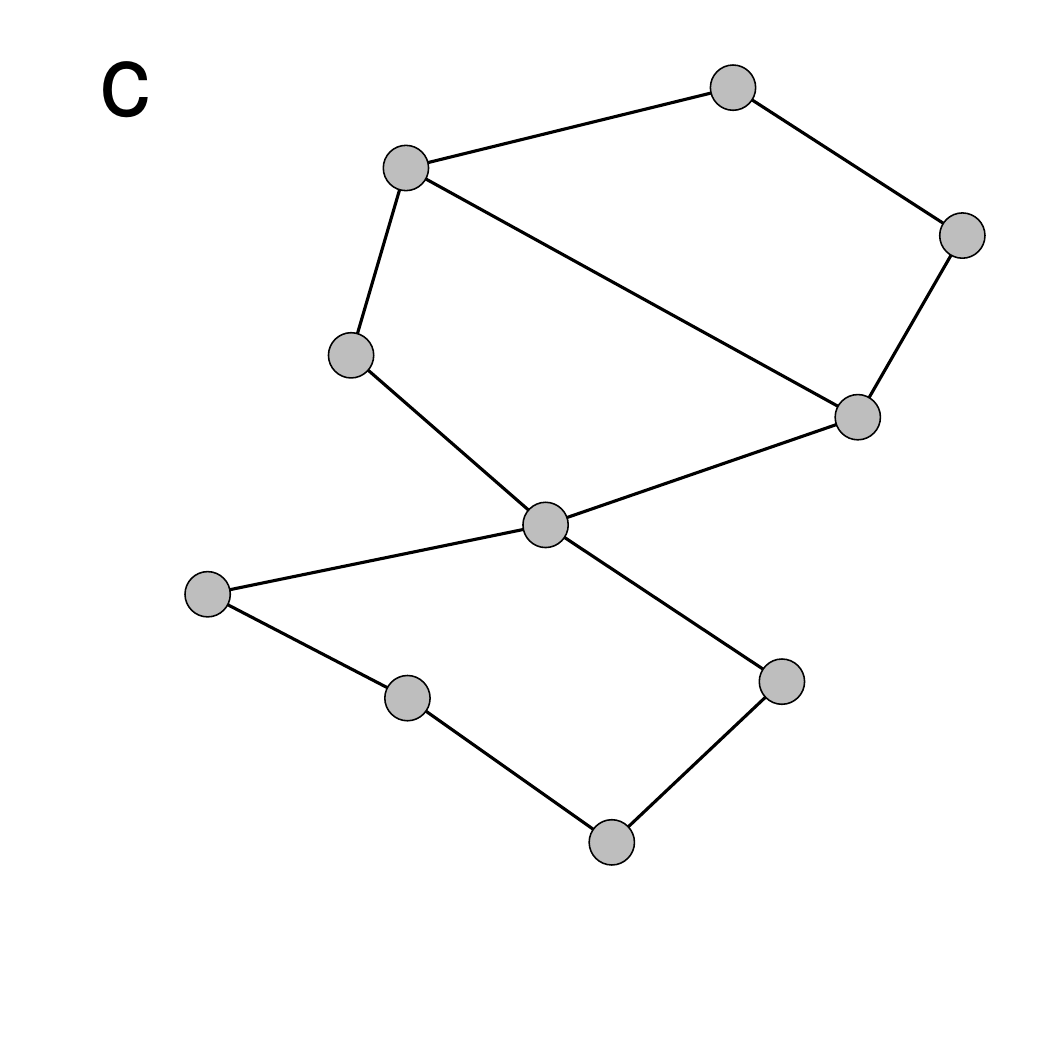}  
\end{center}
\caption{Cycles and chords. ({\bf a}) A triad (smallest cycle) embedded in a network ($\lambda_2 = 0.382$). ({\bf b}) The triadic tie from network {\bf a} is relayed to cross-connect the largest cycle in a way that reduces the mean distance ($\lambda_2 = 0.430$). ({\bf c})  The triadic tie from network {\bf a} is relayed in a way that increases the mean distance ($\lambda_2 = 0.329$).}
\label{fig:cycle} 
\end{figure}

\textit{Chords}. To demonstrate that cross-connecting non-overlapping paths by chords increases $\lambda_2$, we must keep density constant and refrain from disconnecting nodes from the network. These constraints mean that our proof cannot be applied to all graphs (e.g., trees). We use Eq.~\ref{eq:distance2} as our guiding principle. Note that two non-overlapping paths from a sender to a receiver are identical to a cycle wherein the two nodes are embedded. Therefore we can use the fact that for cycles with a length $l \geq 6$, cross-connecting them midway (or almost, when its number of nodes is odd) decreases their mean distance, and more so for longer cycles (Fig.~\ref{fig:distance2}d). For a given network, we take the smallest cycle, usually a triad, we remove one tie from it, and we set it aside. Consequently, this triadic cycle ceases to be without disconnecting nodes, or it becomes a longer cycle; compare Fig.~\ref{fig:cycle}a and b. Either way, we take the longest chordless cycle in the network ($l \geq 6$) and cross-connect it midway with the tie just removed. In isolated cycles, it does not matter where we do this, but in embedded cycles it does matter (Fig.~\ref{fig:cycle}b versus c), hence we carefully place the tie where it reduces the mean distance. Consequently, $\lambda_2$ increases. If we place it awkwardly, $\lambda_2$ decreases (Fig.~\ref{fig:cycle}c). 

\newpage
\section*{Appendix}
\subsection*{Memory convergence}

\begin{figure}[!ht]
\captionsetup{width=.905\linewidth}
\begin{center}
\includegraphics[width=0.35\textwidth]{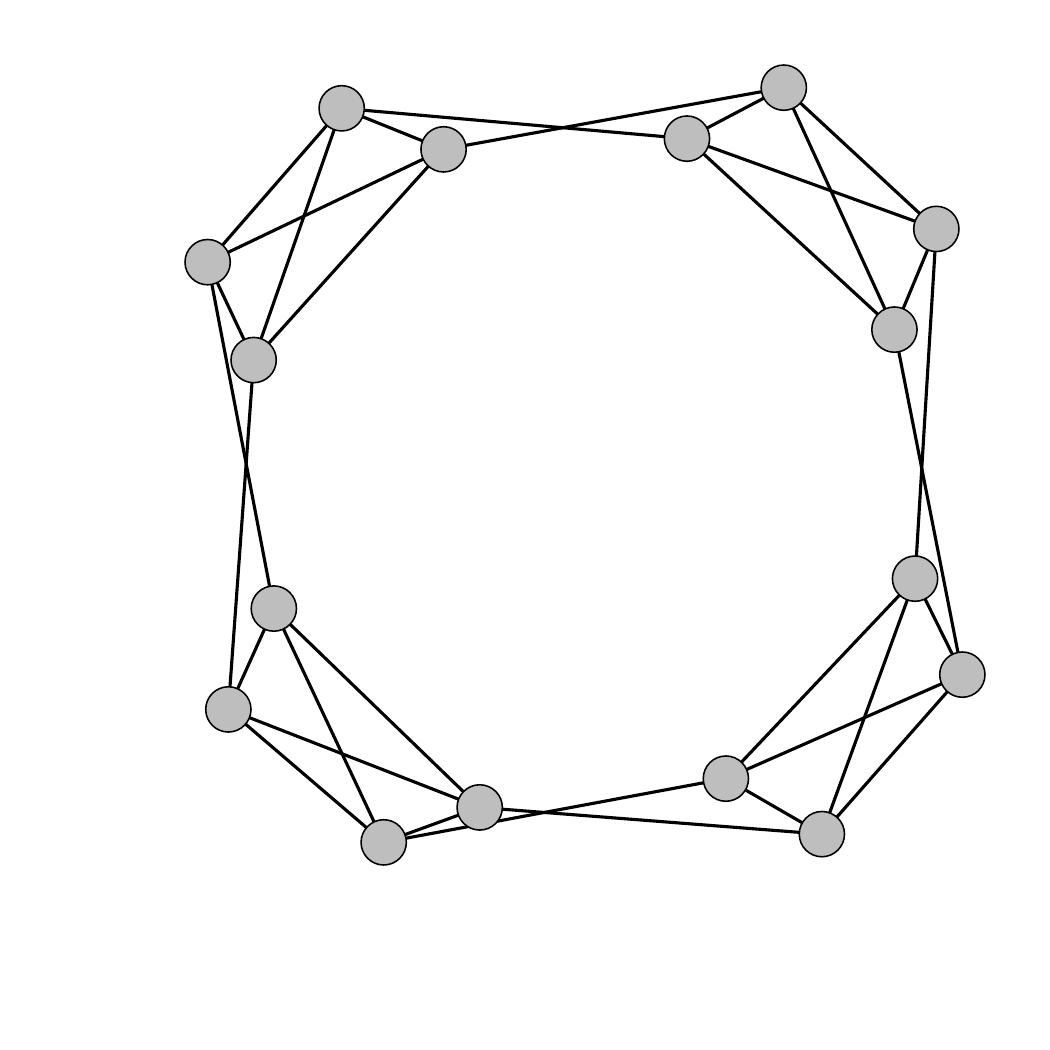} 
\end{center}
\caption{Network with four clusters, used in a mnemonic convergence experiment (adapted from Momennejad et al.~2019).}
\label{fig:momennejad} 
\end{figure}

In a variation of the collective memory experiment, different ties were used over time \cite{momennejad19}. In this variant, there were four clusters with four subjects in each, with connections between the clusters (Fig.~\ref{fig:momennejad}). 
There were two treatments, each with four rounds. In each round, every subject communicated with one other subject. In the first round of treatment~1, subjects communicated between the clusters, not within. In the remaining rounds, they communicated within the clusters, in each round with someone else. In treatment~2, the inter-cluster communication took place in the last round instead of the first. 

Although algebraic connectivity cannot distinguish between the two treatments (because the same network is used in both), the diffusion model (Eq.~1, main text), solved numerically over subsequent rounds, could be applied. 
For the vector of initial memories, the numbers 0 or 1 were assigned randomly to each network node with equal probability. The output of the previous round (through Eq.~1) became the input vector for the next round. The numerical experiment was repeated 1000 times, and averaged. 
If intercluster communication preceded within-cluster communication (treatment~1), there was more convergence of memories than if the sequence was reversed (treatment~2). 
Because more convergence implies a lower standard deviation of memories, the result of the lab experiment can be rephrased as $sd(treatment~2) - sd(treatment~1) > 0$. Consistent with the experiment, the numerical outcome was $sd(treatment~2) - sd(treatment~1) = 0.0297$. 

\subsection*{Data and code availability}
The R code, including all data, is available at \texttt{https://osf.io/8wpxt/}

\small
\bibliographystyle{plain}

\end{document}